\documentclass[pra]{revtex4}
 \usepackage{amssymb} \usepackage{graphicx}

\begin{document}
 \title{Gauged twistor spinors and symmetry operators}

\author{\"Umit Ertem}
 \email{umitertemm@gmail.com}
\address{Department of Physics,
Ankara University, Faculty of Sciences, 06100, Tando\u gan-Ankara,
Turkey\\}

\begin{abstract}

We consider gauged twistor spinors which are supersymmetry generators of supersymmetric and superconformal field theories in curved backgrounds. We show that the spinor bilinears of gauged twistor spinors satify the gauged conformal Killing-Yano equation. We prove that the symmetry operators of the gauged twistor spinor equation can be constructed from ordinary conformal Killing-Yano forms in constant curvature backgrounds. This provides a way to obtain gauged twistor spinors from ordinary twistor spinors.

\end{abstract}

\maketitle

\section{Introduction}

Supersymmetric field theories in flat backgrounds can be generalized to curved backgrounds by coupling them with supergravity and taking the gravity multiplets non-dynamic \cite{Festuccia Seiberg}. Preserved supersymmetries in those cases are found by calculating the vanishing condition of the gravitino variation under supersymmetry transformations. Similarly, superconformal field theories can also be extended to curved backgrounds by coupling to conformal supergravity and they are also studied by using holography \cite{Klare Tomasiello Zaffaroni, Hristov Tomasiello Zaffaroni, Cassani Martelli, Cassani Klare Martelli Tomasiello Zaffaroni}. In all of these cases, supersymmetry generators are determined by the same equation which comes from the variation of the gravitino. It is called the gauged twistor equation or charged conformal Killing spinor equation. It is a generalization of the twistor equation written in terms of the gauged covariant derivative and the gauged Dirac operator \cite{de Medeiros1, de Medeiros2, Lischewski}. Because of the existence of the gauge fields, the solutions of the spinor field equations correspond to $\text{Spin}{^c}$ spinors, namely spinor fields are sections of the bundle that is written as a product of spinor bundle and gauge bundle. The solutions of the gauged twistor equation are called gauged twistor spinors and generate the preserved supersymmetries of supersymmetric and superconformal field theories in curved backgrounds.

One of the methods for finding the solutions of an equation is constructing the symmetry operators of it. Symmetry operators take a solution of the equation and give another solution. The set of mutually commuting symmetry operators are used for finding a general solution by using the method of separation of variables \cite{Miller}. Symmetry operators of some basic spinor field equations can be constructed from the hidden symmetries of the background manifold. Hidden symmetries are defined as the antisymmetric generalizations of Killing vector fields and conformal Killing vector fields to higher degree differential forms. For Killing vector fields, those generalizations are called Killing-Yano (KY) forms and for conformal Killing vector fields, they are conformal Killing-Yano (CKY) forms. The symmetry operators of massless and massive Dirac equations are constructed out of CKY forms and KY forms, respectively \cite{Benn Charlton, Benn Kress, Acik Ertem Onder Vercin, Houri Kubiznak Warnick Yasui, Kubiznak Warnick Krtous, Cariglia Krtous Kubiznak, Cariglia}. Similarly, symmetry operators of geometric Killing spinors are written in terms of odd degree KY forms in constant curvature backgrounds \cite{Ertem1}. CKY forms are used in the construction of the symmetry operators of twistor spinors in constant curvature backgrounds and normal CKY forms play the same role in Einstein manifolds \cite{Ertem2}. Moreover, the symmetry operators of Killing and twistor spinors are also used in the definitions of more general structures such as extended Killing superalgebras and extended conformal superalgebras \cite{Ertem1, Ertem2}.

In this paper, we consider the gauged twistor equation and find its integrability conditions in general $n$ dimensions. We write the spinor bilinears of gauged twistor spinors and show that they correspond to gauged CKY forms which are generalizations of CKY forms with respect to a gauged covariant derivative. We propose a symmetry operator for the gauged twistor equation in terms of CKY forms and prove that it satisfies the required conditions of being a symmetry operator in constant curvature manifolds. Since the spinor bilinears of ordinary twistor spinors correspond to CKY forms, this also opens a way to find the solutions of the gauged twistor equation by using ordinary twistor spinors. This provides a way to find the supersymmetry generators of supersymmetric and superconformal field theories in constant curvature backgrounds.

The paper is organized as follows.  We define the gauged twistor equation and find its integrability conditions in Section 2. In Section 3, we construct the spinor bilinears of gauged twistor spinors  and show that they correspond to gauged CKY forms. A symmetry operator of gauged twistor spinors is proposed in Section 4 and it is proved that it satifies the symmetry operator requirements in constant curvature backgrounds. We also show that the symmetry operator can be written in terms of ordinary twistor spinors. Section 5 concludes the paper.

\section{Gauged twistor spinors}

On a manifold $M$ with $\text{Spin}^c$-structure, one can define a bundle of $U(1)$-valued spinors $S\otimes\Sigma$ where $S$ is the spinor bundle and $\Sigma$ is the $U(1)$ bundle. These types of manifolds can be used to model the backgrounds of supersymmetric field theories on curved space-time. Supersymmetric field theories in flat space-time can be extended to curved space-times by coupling with conformal supergravity and then fixing the gravity multiplets. To preserve some amount of supersymmetry in curved backgrounds, one obtains a condition on the supersymmetry parameters that comes from the variation of the gravitino. The supersymmetry parameters must satisfy the following gauged twistor (or charged conformal Killing spinor) equation in $n$ dimensions;
\begin{equation}
\widehat{\nabla}_X\psi=\frac{1}{n}\widetilde{X}.\widehat{\displaystyle{\not}D}\psi
\end{equation}
with respect to any vector field $X$ and its metric dual $\widetilde{X}$ where $\psi$ is a $\text{Spin}^c$ (or $U(1)$-valued) spinor. Gauged spinor covariant derivative $\widehat{\nabla}_X$ with respect to $X$ is defined in terms of the spinor covariant derivative $\nabla_X$ and gauge connection 1-form $A$, which is generally complex, as
\begin{equation}
\widehat{\nabla}_X:=\nabla_X+i_XA
\end{equation}
where $i_X$ is the interior derivative or contraction operation with respect to $X$. The Dirac operator $\displaystyle{\not}D$ is defined from spinor covariant derivative, frame basis ${X_a}$ and co-frame basis ${e^a}$ with the property $e^a(X_b)=\delta^a_b$ as $\displaystyle{\not}D=e^a.\nabla_{X_a}$ where $.$ denotes the Clifford product. So, the gauged Dirac operator $\widehat{\displaystyle{\not}D}$ in (1) is written as follows
\begin{eqnarray}
\widehat{\displaystyle{\not}D}:=e^a.\widehat{\nabla}_{X_a}=\displaystyle{\not}D+A
\end{eqnarray}
where we have used the expansion of Clifford product in terms of wedge product and interior derivative as $x.\alpha=x\wedge\alpha+i_{\widetilde{x}}\alpha$ for any 1-form $x$, its metric dual $\widetilde{x}$ and any differential $p$-form $\alpha$. We have also used the property $e^a\wedge i_{X_a}\alpha=p\alpha$ \cite{Benn Tucker}.

The exterior derivative $d$ and co-derivative $\delta$ can be written in terms of covariant derivative (with vanishing torsion) as
\begin{equation}
d=e^a\wedge\nabla_{X_a}\quad\quad,\quad\quad\delta=-i_{X^a}\nabla_{X_a}
\end{equation}
and gauged exterior derivative $\widehat{d}$ and co-derivative $\widehat{\delta}$ can be written in terms of them
\begin{eqnarray}
\widehat{d}&:=&e^a\wedge\widehat{\nabla}_{X_a}=d+A\wedge\\
\widehat{\delta}&:=&-i_{X^a}\widehat{\nabla}_{X_a}=\delta-i_{\widetilde{A}}
\end{eqnarray}
where $\widetilde{A}$ is the vector field that is the metric dual of the 1-form $A$. However, on the contrary to the case of $d$ and $\delta$ which satisfy $d^2=\delta^2=0$, the squares of gauged exterior and co-derivatives are written in the following form
\begin{eqnarray}
\widehat{d}^2&=&F\wedge\\
\widehat{\delta}^2&=&-(i_{X^a}i_{X^b}F)i_{X_a}i_{X_b}
\end{eqnarray}
where $F=dA$ is the curvature of the gauge connection 1-form $A$ \cite{Charlton}.

\subsection{Integrability conditions}

The existence of gauged twistor spinors in a manifold depends on some integrability conditions of (1) which constrain the curvature characteristics of the background manifold. They can be obtained by taking second covariant derivatives of the gauged twistor equation and by using the following definition of the curvature operator of the gauged covariant derivative
\begin{equation}
\widehat{R}(X,Y)=[\widehat{\nabla}_X,\widehat{\nabla}_Y]-\widehat{\nabla}_{[X,Y]}
\end{equation}
where $X$ and $Y$ are arbitrary vector fields. From the definition (2), it can be written in terms of the curvature operator $R(X,Y)$ of the Levi-Civita connection and the curvature $F$ of the gauge connection as follows
\begin{equation}
\widehat{R}(X_a,X_b)=R(X_a,X_b)+i_{X_b}i_{X_a}F
\end{equation}
where $\{X_a\}$ is an orthonormal frame. The action of the curvature operator $R(X_a,X_b)$ on a spinor $\psi$ can be written in terms of curvature 2-forms $R_{ab}$ as $R(X_a, X_b)\psi=\frac{1}{2}R_{ab}.\psi$ \cite{Benn Tucker, Charlton}. From the curvature 2-forms $R_{ab}$, the definition of Ricci 1-forms $P_a$ and curvature scalar ${\cal{R}}$ can be stated as $P_a=i_{X^b}R_{ba}$ and ${\cal{R}}=i_{X^a}P_a$, respectively.

The action of the operator in (10) on a gauged twistor spinor $\psi$ must be equal to the action of the right hand side of (9) on the same gauged twistor spinor. By using (1), we obtain
\begin{equation}
R(X_a,X_b)\psi+(i_{X_b}i_{X_a}F)\psi=\frac{1}{n}\widehat{\nabla}_{X_a}(e_b.\widehat{\displaystyle{\not}D}\psi)-\frac{1}{n}\widehat{\nabla}_{X_b}(e_a.\widehat{\displaystyle{\not}D}\psi).
\end{equation} 
So, one can write the action of curvature 2-forms $R_{ab}$ on gauged twistor spinors as follows
\begin{equation}
R_{ab}.\psi=\frac{2}{n}\left(e_b.\widehat{\nabla}_{X_a}\widehat{\displaystyle{\not}D}\psi-e_a.\widehat{\nabla}_{X_b}\widehat{\displaystyle{\not}D}\psi\right)-2(i_{X_b}i_{X_a}F)\psi.
\end{equation}
For zero torsion, we have the equalities $R_{ab}\wedge e^a=0$ and $e^a.R_{ab}=P_b$. By using them, the action of Ricci 1-forms $P_a$ on gauged twistor spinors can be calculated from (12)
\begin{eqnarray}
P_b.\psi&=&\frac{2}{n}\left(e^a.e_b.\widehat{\nabla}_{X_a}\widehat{\displaystyle{\not}D}\psi-e^a.e_a.\widehat{\nabla}_{X_b}\widehat{\displaystyle{\not}D}\psi\right)-2e^a.(i_{X_b}i_{X_a}F)\psi\nonumber\\
&=&-\frac{2}{n}e_b.\widehat{\displaystyle{\not}D}^2\psi-\frac{2(n-2)}{n}\widehat{\nabla}_{X_b}\widehat{\displaystyle{\not}D}\psi+2(i_{X_b}F).\psi
\end{eqnarray}
where we have used the Clifford algebra identity $e^a.e_b+e_b.e^a=2g^a_b$ for the metric $g_{ab}$ and the definition (3). Similarly, Ricci 1-forms satisfy the equalities $P_a\wedge e^a=0$ and $e^a.P_a={\cal{R}}$. So, we can write the action of the curvature scalar ${\cal{R}}$ on gauged twistor spinors from (13)
\begin{eqnarray}
{\cal{R}}\psi&=&-\frac{2}{n}e^a.e_a.\widehat{\displaystyle{\not}D}^2\psi-\frac{2(n-2)}{n}e^a.\widehat{\nabla}_{X_a}\widehat{\displaystyle{\not}D}\psi+2e^a.i_{X_a}F.\psi\nonumber\\
&=&-\frac{4(n-1)}{n}\widehat{\displaystyle{\not}D}^2\psi+4F.\psi.
\end{eqnarray}
By combining (13) and (14), one can obtain the following two integrability conditions of the gauged twistor equation
\begin{equation}
\widehat{\displaystyle{\not}D}^2\psi=-\frac{n}{4(n-1)}{\cal{R}}\psi+\frac{n}{n-1}F.\psi
\end{equation}
\begin{equation}
\widehat{\nabla}_{X_a}\widehat{\displaystyle{\not}D}\psi=\frac{n}{2}K_a.\psi-\frac{n}{(n-1)(n-2)}e_a.F.\psi+\frac{n}{n-2}i_{X_a}F.\psi
\end{equation}
where the 1-form $K_a$ is defined as follows
\begin{equation}
K_a=\frac{1}{n-2}\left(\frac{\cal{R}}{2(n-1)}e_a-P_a\right).
\end{equation}
Moreover, from the definition of the conformal 2-forms
\begin{equation}
C_{ab}=R_{ab}-\frac{1}{n-2}\left(P_a\wedge e_b-P_b\wedge e_a\right)+\frac{1}{(n-1)(n-2)}{\cal{R}}e_{ab}
\end{equation}
where $e_{ab}=e_a\wedge e_b$, the third integrability condition that corresponds to the action of $C_{ab}$ on gauged twistor spinors can be found from (12), (13) and (14) as
\begin{equation}
C_{ab}.\psi=2(i_{X_a}i_{X_b}F)\psi+\frac{2}{n-2}\left(e_b.i_{X_a}F-e_a.i_{X_b}F\right).\psi+\frac{4}{(n-1)(n-2)}e_a.e_b.F.\psi.
\end{equation}
For $A=0$, (15), (16) and (19) reduce to the integrability conditions of the ordinary twistor spinor equation \cite{Baum Friedrich Grunewald Kath, Baum Leitner, Benn Kress2}.

Besides being necessary conditions on gauged twistor spinors, the equalities (15), (16) and (19) also determine the existence conditions for gauged twistors. For the Spin$^c$ bundle $S'$, by defining the bundle $E=S'\oplus S'$ and obtaining the curvature operator of the bundle, one can see that the action of the curvature operator on $(\psi, \displaystyle{\not}{\widehat{D}}\psi)$ vanishes for a gauged twistor spinor $\psi$, becasue of (15), (16) and (19) \cite{Lischewski}. From this result, a partial classification for manifolds admitting gauged twistor spinors can be determined. For example, Lorentzian Einstein-Sasaki manifolds, Fefferman spaces and a product of Lorentizan Einsten-Sasaki manifolds and Riemannnian Einstein manifolds can admit gauged twistor spinors \cite{Lischewski}.

\section{Spinor bilinears and gauged CKY forms}

The tensor product of the spinor space $S$ and dual spinor space $S^*$ correspond to the algebra of endomorphisms over the spinor space; $S\otimes S^*=\text{End}(S)$ and it is isomorphic to the Clifford algebra of the relevant dimension which is also isomorphic to the exteriror algebra $\Lambda M$ of differential forms on $M$. So, the tensor products of spinors and its duals which are called spinor bilinears can be written as a sum of different degree differential forms
\begin{equation}
\psi\otimes\overline{\psi}=(\psi,\psi)+(\psi, e_a.\psi)e^a+(\psi, e_{ba}.\psi)e^{ab}+...+(\psi, e_{a_p...a_2a_1}.\psi)e^{a_1a_2...a_p}+...+(-1)^{\lfloor{n/2}\rfloor}(\psi, z.\psi)z
\end{equation}
where $e^{a_1a_2...a_p}=e^{a_1}\wedge e^{a_2}\wedge...\wedge e^{a_p}$, $\lfloor{ }\rfloor$ is the floor function that takes the integer part of the argument,   $z$ is the volume form and $(\,,\,)$ denotes the spinor inner product. Every $p$-form component on the right hand side of (20) is called the $p$-form Dirac current as the generalization of the Dirac current that corresponds to the metric dual of the 1-form part of the spinor bilinear \cite{Acik Ertem}. $p$-form Dirac currents will be denoted as follows
\begin{equation}
(\psi\overline{\psi})_p=(\psi, e_{a_p...a_2a_1}.\psi)e^{a_1a_2...a_p}.
\end{equation}
For a gagued twistor spinor $\psi$, by requiring that the connection $\widehat{\nabla}$ is compatible with the spinor inner product $(\,,\,)$, we will show that the $p$-form Dirac currents of gauged twistor spinors satisfy the gauged CKY equation which is the generalization of the CKY equation that corresponds to the antisymmetric generalization of the conformal Killing equation to higher degree forms.

After applying the gauged covariant derivative to (21) and doing some manipulations, one obtains that
\begin{eqnarray}
\widehat{\nabla}_{X_a}(\psi\overline{\psi})_p&=&\left((\widehat{\nabla}_{X_a}\psi)\overline{\psi}\right)_p+\left(\psi\overline{\widehat{\nabla}_{X_a}\psi}\right)_p\nonumber\\
&=&\frac{1}{n}\left((e_a.\widehat{\displaystyle{\not}D}\psi)\overline{\psi}\right)_p+\frac{1}{n}\left(\psi\overline{e_a.\widehat{\displaystyle{\not}D}\psi}\right)_p\nonumber\\
&=&\frac{1}{n}\left(e_a.\widehat{\displaystyle{\not}d}(\psi\overline{\psi})\right)_p-\frac{1}{n}\left(e_a.e_b.\psi\overline{\widehat{\nabla}_{X_b}\psi}\right)_p+\frac{1}{n}\left(\psi\overline{\widehat{\nabla}_{X_b}\psi}.e_b.e_a\right)_p\nonumber
\end{eqnarray}
where we have used (1) and $(\widehat{\nabla}_{X_a}\psi)\overline{\psi}=\widehat{\nabla}_{X_a}(\psi\overline{\psi})-\psi(\overline{\widehat{\nabla}_{X_a}\psi})$ with the definition $\widehat{\displaystyle{\not}d}=e_a.\widehat{\nabla}_{X^a}$ on differential forms. From (5) and (6), one can write $\widehat{\displaystyle{\not}d}=\widehat{d}-\widehat{\delta}$ and obtain the following equality by using the definition of the Clifford product of a 1-form with any $p$-form in terms of the wedge product and interior derivative
\begin{eqnarray}
\widehat{\nabla}_{X_a}(\psi\overline{\psi})_p&=&\frac{1}{n}\left(e_a\wedge\widehat{d}(\psi\overline{\psi})_{p-2}+i_{X_a}\widehat{d}(\psi\overline{\psi})_p\right)-\frac{1}{n}\left(e_a\wedge(e_b.\psi\overline{\widehat{\nabla}_{X^b}\psi})_{p-1}+i_{X_a}(e_b.\psi\overline{\widehat{\nabla}_{X^b}\psi})_{p+1}\right)\nonumber\\
&&-\frac{1}{n}\left(e_a\wedge\widehat{\delta}(\psi\overline{\psi})_p+i_{X_a}\widehat{\delta}(\psi\overline{\psi})_{p+2}\right)\pm\frac{1}{n}\left(e_a\wedge(\psi\overline{\widehat{\nabla}_{X^b}\psi}.e_b)_{p-1}+i_{X_a}(\psi\overline{\widehat{\nabla}_{X^b}\psi}.e_b)_{p+1}\right)\nonumber\\
\end{eqnarray}
where $\pm$ sign depends on the chosen inner automorphism of the Clifford algebra which is used in the definition of the duality operation $\bar{\quad}$. By wedge multiplying (22) with $e^a$ from the left and using (5), we can write
\begin{equation}
\widehat{d}(\psi\overline{\psi})_p=\frac{p+1}{n}\left(\widehat{d}(\psi\overline{\psi})_p-\widehat{\delta}(\psi\overline{\psi})_{p+2}\right)-\frac{p+1}{n}\left((e_b.\psi\overline{\widehat{\nabla}_{X^b}\psi})_{p+1}\mp(\psi\overline{\widehat{\nabla}_{X^b}\psi}.e_b)_{p+1}\right)
\end{equation}
and similarly by taking the interior derivative of (22) with respect to $X_a$ and using (6), it can also be written
\begin{equation}
\widehat{\delta}(\psi\overline{\psi})_p=-\frac{n-p+1}{n}\left(\widehat{d}(\psi\overline{\psi})_{p-2}-\widehat{\delta}(\psi\overline{\psi})_p\right)+\frac{n-p+1}{n}\left((e_b.\psi\overline{\widehat{\nabla}_{X^b}\psi})_{p-1}\mp(\psi\overline{\widehat{\nabla}_{X^b}\psi}.e_b)_{p-1}\right).
\end{equation}
So, by comparing (22), (23) and (24), one can see that the $p$-form Dirac currents of gauged twistor spinors satisfy the following equation
\begin{equation}
\widehat{\nabla}_{X_a}(\psi\overline{\psi})_p=\frac{1}{p+1}i_{X_a}\widehat{d}(\psi\overline{\psi})_p-\frac{1}{n-p+1}e_a\wedge\widehat{\delta}(\psi\overline{\psi})_p.
\end{equation}
This equation is called the gauged CKY equation. In general, a $p$-form $\omega$ is called a gauged CKY $p$-form, if it satisfies the following gauged CKY equation
\begin{equation}
\widehat{\nabla}_{X_a}\omega=\frac{1}{p+1}i_{X_a}\widehat{d}\omega-\frac{1}{n-p+1}e_a\wedge\widehat{\delta}\omega.
\end{equation}
From (2), it can also be written in terms of the Levi-Civita connection $\nabla$ and the gauge potential 1-form $A$ as
\begin{eqnarray}
\nabla_{X_a}\omega-\frac{1}{p+1}i_{X_a}d\omega+\frac{1}{n-p+1}e_a\wedge\delta\omega\nonumber\\
=-\frac{p}{p+1}(i_{X_a}A)\omega-\frac{1}{p+1}A\wedge i_{X_a}\omega+\frac{1}{n-p+1}e_a\wedge i_{\widetilde{A}}\omega.
\end{eqnarray}
For $A=0$, it reduces to the ordinary CKY equation which is the antisymmetric generalization of the conformal Killing equation to higher degree forms
\begin{equation}
\nabla_{X_a}\omega=\frac{1}{p+1}i_{X_a}d\omega-\frac{1}{n-p+1}e_a\wedge\delta\omega.
\end{equation}
For $p=1$, (27) reduces to the shear-free vector field equation which is the generalization of the conformal Killing equation and describes the vector fields that constitute shear-free congruences \cite{Charlton}.

Integrability conditions of the gauged CKY equation can be calculated by taking second covariant derivatives of (26). After some manipulations, they can be obtained as follows
\begin{eqnarray}
\widehat{\nabla}_{X_b}\widehat{d}\omega&=&\frac{p+1}{p}R_{ab}\wedge i_{X^a}\omega+\frac{p+1}{p(n-p+1)}e_b\wedge\widehat{d}\widehat{\delta}\omega\nonumber\\
&&+i_{X_b}F\wedge\omega-\frac{1}{p}F\wedge i_{X_b}\omega+\frac{p+1}{p(n-p+1)}e_b\wedge A\wedge\widehat{\delta}\omega
\end{eqnarray}
\begin{eqnarray}
\widehat{\nabla}_{X_b}\widehat{\delta}\omega&=&\frac{n-p+1}{n-p}\bigg((i_{X_a}P_b)i_{X^a}\omega+i_{X_a}R_{cb}\wedge i_{X^c}i_{X^a}\omega+(i_{X_b}i_{X_a}F)i_{X^a}\omega\bigg)\nonumber\\
&&-\frac{n-p+1}{(p+1)(n-p)}i_{X_b}\widehat{\delta}\widehat{d}\omega-(i_{X_b}A)\widehat{\delta}\omega-\frac{1}{n-p}e_b\wedge\bigg(i_{\widetilde{A}}\widehat{\delta}\omega+(i_{X_a}i_{X_c}F)i_{X^a}i_{X^c}\omega\bigg)\nonumber\\
\end{eqnarray}
and their combination gives
\begin{eqnarray}
\frac{p}{p+1}\widehat{\delta}\widehat{d}\omega+\frac{n-p}{n-p+1}\widehat{d}\widehat{\delta}\omega&=&P_a\wedge i_{X^a}\omega+R_{ab}\wedge i_{X^a}i_{X^b}\omega\nonumber\\
&&-\frac{n-p}{n-p+1}A\wedge\widehat{\delta}\omega+i_{X^a}F\wedge i_{X_a}\omega.
\end{eqnarray}
For $A=0$, they reduce to the integrability conditions of the ordinary CKY equation which read as \cite{Semmelmann, Ertem3}
\begin{equation}
\nabla_{X_b}d\omega=\frac{p+1}{p}R_{ab}\wedge i_{X^a}\omega+\frac{p+1}{p(n-p+1)}e_b\wedge d\delta\omega
\end{equation}
\begin{equation}
\nabla_{X_b}\delta\omega=\frac{n-p+1}{n-p}\bigg((i_{X_a}P_b)i_{X^a}\omega+i_{X_a}R_{cb}\wedge i_{X^c}i_{X^a}\omega\bigg)-\frac{n-p+1}{(p+1)(n-p)}i_{X_b}\delta d\omega
\end{equation}
\begin{equation}
\frac{p}{p+1}\delta d\omega+\frac{n-p}{n-p+1}d\delta\omega=P_a\wedge i_{X^a}\omega+R_{ab}\wedge i_{X^a}i_{X^b}\omega.
\end{equation}

\section{Symmetry operators}

Solutions of the gauged twistor equation (1) gives the supersymmetry parameters of supersymmetric field theories coupled with conformal supergravity. So, finding a solution generating technique for (1) is an important problem. Rather than solving an equation directly, one can also construct symmetry operators of it which give the solutions of an equation from a known solution. For example, symmetry operators of massless and massive Dirac equation can be constructed from CKY and KY forms, respectively \cite{Benn Charlton, Benn Kress}. Similarly, symmetry operators of geometric Killing spinors and ordinary twistor spinors can also be written in terms of KY and CKY forms respectively in constant curvature manifolds \cite{Ertem1, Ertem2}. We can search for the symmetry operators of the gauged twistor equation in terms of gauged or ordinary CKY forms. We propose the following operator
\begin{eqnarray}
L_{\omega}&=&-(-1)^p\frac{p}{n}\omega.\widehat{\displaystyle{\not}D}+\frac{p}{2(p+1)}d\omega+\frac{p}{2(n-p+1)}\delta\omega\nonumber\\
&=&-(-1)^p\frac{p}{n}\omega.\displaystyle{\not}D+\frac{p}{2(p+1)}d\omega+\frac{p}{2(n-p+1)}\delta\omega-(-1)^p\frac{p}{n}\omega.A
\end{eqnarray}
written in terms of ordinary CKY $p$-forms $\omega$. Note that $d$ and $\delta$ are exterior and co-derivatives with respect to the Levi-Civita connection and $\widehat{\displaystyle{\not}D}$ is the gauged Dirac operator (3). Eq. (35) reduces to the symmetry operators of ordinary twistor spinors for $A=0$.

To prove that (35) is a symmetry operator for the gauged twistor equation, we need to show that if $\psi$ is a gauged twistor spinor, then $L_{\omega}\psi$ is a solution of the gauged twistor equation, namely it satisfies the following equality
\begin{equation}
\widehat{\nabla}_{X_a}L_{\omega}\psi=\frac{1}{n}e_a.\widehat{\displaystyle{\not}D}L_{\omega}\psi
\end{equation}
which can also be written in terms the Levi-Civita connection as
\begin{equation}
\nabla_{X_a}L_{\omega}\psi-\frac{1}{n}e_a.\displaystyle{\not}DL_{\omega}\psi=-\frac{n-2}{2n}e_a.A.L_{\omega}\psi-\frac{1}{2}A.e_a.L_{\omega}\psi.
\end{equation}
So, we will expand all the terms in (37) to check the equality. By using (35), the first term on the left hand side of (37) can be obtained as follows
\begin{eqnarray}
\nabla_{X_a}L_{\omega}\psi&=&-(-1)^p\frac{p}{n}\nabla_{X_a}\omega.\widehat{\displaystyle{\not}D}\psi-(-1)^p\frac{p}{n}\omega.\nabla_{X_a}\widehat{\displaystyle{\not}D}\psi+\frac{p}{2(p+1)}\nabla_{X_a}d\omega.\psi\nonumber\\
&&+\frac{p}{2(p+1)}d\omega.\nabla_{X_a}\psi+\frac{p}{2(n-p+1)}\nabla_{X_a}\delta\omega.\psi+\frac{p}{2(n-p+1)}\delta\omega.\nabla_{X_a}\psi.
\end{eqnarray}
Here, we can use (1) and (16) which are written in terms of the Levi-Civita connection as
\begin{equation}
\nabla_{X_a}\psi=\frac{1}{n}e_a.\widehat{\displaystyle{\not}D}\psi-\frac{1}{2}(e_a.A+A.e_a).\psi
\end{equation}
\begin{equation}
\nabla_{X_a}\widehat{\displaystyle{\not}D}\psi=\frac{n}{2}K_a.\psi-\frac{n}{(n-1)(n-2)}e_a.F.\psi+\frac{n}{n-2}i_{X_a}F.\psi-\frac{1}{2}(e_a.A+A.e_a).\widehat{\displaystyle{\not}D}\psi.
\end{equation}
Hence, (38) transforms into the following form
\begin{eqnarray}
\nabla_{X_a}L_{\omega}\psi&=&\bigg[-(-1)^p\frac{p}{n}\nabla_{X_a}\omega+(-1)^p\frac{p}{2n}(e_a.A+A.e_a).\omega\nonumber\\
&&+\frac{p}{2n(p+1)}d\omega.e_a+\frac{p}{2n(n-p+1)}\delta\omega.e_a\bigg].\widehat{\displaystyle{\not}D}\psi\nonumber\\
&&+\bigg[-(-1)^p\frac{p}{2}\omega.K_a+(-1)^p\frac{p}{(n-1)(n-2)}\omega.e_a.F-(-1)^p\frac{p}{n-2}\omega.i_{X_a}F\nonumber\\
&&+\frac{p}{2(p+1)}\nabla_{X_a}d\omega-\frac{p}{4(p+1)}(e_a.A+A.e_a).d\omega+\frac{p}{2(n-p+1)}\nabla_{X_a}\delta\omega\nonumber\\
&&-\frac{p}{4(n-p+1)}(e_a.A+A.e_a).\delta\omega\bigg].\psi
\end{eqnarray}
where we have used the fact that $e_a.A+A.e_a=2i_{X_a}A$ is a function and it commutes with the differential forms $\omega$, $d\omega$ and $\delta\omega$. The second term on the left hand side of (37) can also be written from (41) and we obtain
\begin{eqnarray}
-\frac{1}{n}e_a.\displaystyle{\not}DL_{\omega}\psi&=&-\frac{1}{n}e_a.e^b.\nabla_{X_b}L_{\omega}\psi\nonumber\\
&=&\bigg[(-1)^p\frac{p}{n^2}e_a.e^b.\nabla_{X_b}\omega-(-1)^p\frac{p}{n^2}e_a.A.\omega\nonumber\\
&&-\frac{p}{2n^2(p+1)}e_a.e^b.d\omega.e_b-\frac{p}{2n^2(n-p+1)}e_a.e^b.\delta\omega.e_b\bigg].\widehat{\displaystyle{\not}D}\psi\nonumber\\
&&+\bigg[(-1)^p\frac{p}{2n}e_a.e^b.\omega.K_b-(-1)^p\frac{p}{n(n-1)(n-2)}e_a.e^b.\omega.e_b.F\nonumber\\
&&+(-1)^p\frac{p}{n(n-2)}e_a.e^b.\omega.i_{X_b}F-\frac{p}{2n(p+1)}e_a.e^b.\nabla_{X_b}d\omega++\frac{p}{2n(p+1)}e_a.A.d\omega\nonumber\\
&&-\frac{p}{2n(n-p+1)}e_a.e^b.\nabla_{X_b}\delta\omega+\frac{p}{2n(n-p+1)}e_a.A.\delta\omega\bigg].\psi\nonumber\\
\end{eqnarray}
where we have simplified the terms by using again the relation $e_a.A+A.e_a=2i_{X_a}A$ and $A=(i_{X_a}A)e^a$. Similarly, by using (35), we can write the terms on the right hand side of (37) in the following way
\begin{eqnarray}
-\frac{n-2}{2n}e_a.A.L_{\omega}\psi&=&(-1)^p\frac{p(n-2)}{2n^2}e_a.A.\omega.\widehat{\displaystyle{\not}D}\psi\nonumber\\
&&-\bigg[\frac{p(n-2)}{4n(p+1)}e_a.A.d\omega+\frac{p(n-2)}{4n(n-p+1)}e_a.A.\delta\omega\bigg].\psi
\end{eqnarray}
and
\begin{eqnarray}
-\frac{1}{2}A.e_a.L_{\omega}\psi&=&(-1)^p\frac{p}{2n}A.e_a.\omega.\widehat{\displaystyle{\not}D}\psi\nonumber\\
&&-\bigg[\frac{p}{4(p+1)}A.e_a.d\omega+\frac{p}{4(n-p+1)}A.e_a.\delta\omega\bigg].\psi.
\end{eqnarray}

Now, we write all the terms in (37) explicitly and we are in a position to check the correctness of (37) by comparing the equalities in (41)-(44). We will do this in two steps, since we can consider the coefficients of $\widehat{\displaystyle{\not}D}\psi$ and $\psi$ separately in each equality. So, as a first step, the terms in the coefficients of $\widehat{\displaystyle{\not}D}\psi$ for the terms on the left hand side of (37) must be equal to the coefficients of $\widehat{\displaystyle{\not}D}\psi$ for the terms on the right hand side of (37). We know that $\omega$ is an ordinary CKY $p$-form and satisfies (28), so the sum of the coefficients of $\widehat{\displaystyle{\not}D}\psi$ in (41) and (42) (corresponding to the left hand side of (37)) can  be written as
\begin{eqnarray}
&&-(-1)^p\frac{p}{n(p+1)}i_{X_a}d\omega+(-1)^p\frac{p}{n(n-p+1)}e_a\wedge\delta\omega+\frac{p}{2n(p+1)}d\omega.e_a+\frac{p}{2n(n-p+1)}\delta\omega.e_a\nonumber\\
&&+(-1)^p\frac{p}{n^2(p+1)}e_a.e^b.i_{X_b}d\omega-(-1)^p\frac{p}{n^2(n-p+1)}e_a.e^b.(e_b\wedge\delta\omega)\nonumber\\
&&-\frac{p}{2n^2(p+1)}e_a.e^b.d\omega.e_b-\frac{p}{2n^2(n-p+1)}e_a.e^b.\delta\omega.e_b\nonumber\\
&&+(-1)^p\frac{p}{2n}(e_a.A+A.e_a).\omega-(-1)^p\frac{p}{n^2}e_a.A.\omega\nonumber\\
&=&(-1)^p\frac{p}{2n}(e_a.A+A.e_a).\omega-(-1)^p\frac{p}{n^2}e_a.A.\omega
\end{eqnarray}
where we have used the expansion of the Clifford product in terms of the wedge product and interior derivative as follows
\begin{eqnarray}
d\omega.e_a&=&-(-1)^pe_a\wedge d\omega+(-1)^pi_{X_a}d\omega,\nonumber\\
\delta\omega.e_a&=&-(-1)^pe_a\wedge\delta\omega+(-1)^pi_{X_a}\delta\omega
\end{eqnarray}
and from the equality $e^a.\omega.e_a=(-1)^p(n-2p)\omega$
\begin{eqnarray}
e_a.e^b.i_{X_b}d\omega&=&(p+1)e_a\wedge d\omega+(p+1)i_{X_a}d\omega,\nonumber\\
e_a.e^b.(e_b\wedge\delta\omega)&=&(n-p+1)e_a\wedge\delta\omega+(n-p+1)i_{X_a}\delta\omega,\nonumber\\
e_a.e^b.d\omega.e_b&=&-(-1)^p(n-2(p+1))e_a\wedge d\omega-(-1)^p(n-2(p+1))i_{X_a}d\omega,\\
e_a.e^b.\delta\omega.e_b&=&-(-1)^p(n-2(p-1))e_a\wedge\delta\omega-(-1)^p(n-2(p-1))i_{X_a}\delta\omega.\nonumber
\end{eqnarray}
So, the terms that do not contain $A$ on the left hand side of (45) cancel each other and we obtain the right hand side of (45). On the other hand, one can easily see that, for the sum of the coefficients of $\widehat{\displaystyle{\not}D}\psi$ in (43) and (44) (corresponding to the right hand side of (37)) is exactly equal to the right hand side of (45). Hence, we prove the first step, that is the coefficients of $\widehat{\displaystyle{\not}D}\psi$ in the equalities (41)-(44) satisfy (37).

As a second step, we will consider the coefficients of $\psi$ for the terms in (37) by using (41)-(44). We can write the coefficients of $\psi$ in (37) in the following form
\begin{eqnarray}
&&-(-1)^p\frac{p}{2}\omega.K_a+\frac{p}{2(p+1)}\nabla_{X_a}d\omega+\frac{p}{2(n-p+1)}\nabla_{X_a}\delta\omega\nonumber\\
&&+(-1)^p\frac{p}{2n}e_a.e^b.\omega.K_b-\frac{p}{2n(p+1)}e_a.e^b.\nabla_{X_b}d\omega-\frac{p}{2n(n-p+1)}e_a.e^b.\nabla_{X_b}\delta\omega\nonumber\\
&&+(-1)^p\frac{p}{(n-1)(n-2)}\omega.e_a.F-(-1)^p\frac{p}{n-2}\omega.i_{X_a}F\nonumber\\
&&-(-1)^p\frac{p}{n(n-1)(n-2)}e_a.e^b.\omega.e_b.F+(-1)^p\frac{p}{n(n-2)}e_a.e^b.\omega.i_{X_b}F\nonumber\\
&&-\frac{p}{4(p+1)}(e_a.A.d\omega+A.e_a.d\omega)-\frac{p}{4(n-p+1)}(e_a.A.\delta\omega+A.e_a.\delta\omega)\nonumber\\
&&+\frac{p}{2n(p+1)}e_a.A.d\omega+\frac{p}{2n(n-p+1)}e_a.A.\delta\omega+\frac{p(n-2)}{4n(p+1)}e_a.A.d\omega\nonumber\\
&&+\frac{p(n-2)}{4n(n-p+1)}e_a.A.\delta\omega+\frac{p}{4(p+1)}A.e_a.d\omega+\frac{p}{4(n-p+1)}A.e_a.\delta\omega\nonumber\\
&=&0
\end{eqnarray}
and we need to check that the left hand side of (48) is equal to zero. As can easily be seen that the terms that contain $A$ on the left hand side of (48) cancel each other and we obtain
\begin{eqnarray}
&&-(-1)^pp\omega.\bigg[\frac{K_a}{2}-\frac{1}{(n-1)(n-2)}e_a.F+\frac{1}{n-2}i_{X_a}F\bigg]\nonumber\\
&&+(-1)^p\frac{p}{n}e_a.e^b.\omega.\bigg[\frac{K_b}{2}-\frac{1}{(n-1)(n-2)}e_b.F+\frac{1}{n-2}i_{X_b}F\bigg]\nonumber\\
&&+\frac{p}{2(p+1)}\bigg[\nabla_{X_a}d\omega-\frac{1}{n}e_a.e^b.\nabla_{X_b}d\omega\bigg]\nonumber\\
&&+\frac{p}{2(n-p+1)}\bigg[\nabla_{X_a}\delta\omega-\frac{1}{n}e_a.e^b.\nabla_{X_b}\delta\omega\bigg]\nonumber\\
&=&0.
\end{eqnarray}
We know that $\omega$ is an ordinary CKY $p$-form and it satisfies the integrability conditions in (32)-(34). Considering this and the following equalities (by using (34))
\begin{eqnarray}
e_a.e^b.\nabla_{X_b}d\omega&=&-e_a\wedge\delta d\omega-i_{X_a}\delta d\omega\nonumber\\
&=&\frac{(p+1)(n-p)}{p(n-p+1)}e_a\wedge d\delta\omega-i_{X_a}\delta d\omega\nonumber\\
&&-\frac{p+1}{p}\bigg[e_a\wedge P_b\wedge i_{X^b}\omega+e_a\wedge R_{bc}\wedge i_{X^b}i_{X^c}\omega\bigg]
\end{eqnarray}
\begin{eqnarray}
e_a.e^b.\nabla_{X_b}\delta\omega&=&e_a\wedge d\delta\omega+i_{X_a}d\delta\omega\nonumber\\
&=&e_a\wedge d\delta\omega-\frac{p(n-p+1)}{(p+1)(n-p)}i_{X_a}\delta d\omega\nonumber\\
&&+\frac{n-p+1}{n-p}\bigg[i_{X_a}(P_b\wedge i_{X^b}\omega)+i_{X_a}(R_{bc}\wedge i_{X^b}i_{X^c}\omega)\bigg]
\end{eqnarray}
(49) transforms into
\begin{eqnarray}
&&-(-1)^pp\omega.\bigg[\frac{K_a}{2}-\frac{1}{(n-1)(n-2)}e_a.F+\frac{1}{n-2}i_{X_a}F\bigg]\nonumber\\
&&+(-1)^p\frac{p}{n}e_a.e^b.\omega.\bigg[\frac{K_b}{2}-\frac{1}{(n-1)(n-2)}e_b.F+\frac{1}{n-2}i_{X_b}F\bigg]\nonumber\\
&&+\frac{1}{2}R_{ba}\wedge i_{X^b}\omega+\frac{1}{2n}\bigg(e_a\wedge P_b\wedge i_{X^b}\omega+e_a\wedge R_{bc}\wedge i_{X^b}i_{X^c}\omega\bigg)\nonumber\\
&&+\frac{p}{2(n-p)}\bigg((i_{X_b}P_a)i_{X^b}\omega+i_{X_b}R_{ca}\wedge i_{X^c}i_{X^b}\omega\bigg)\nonumber\\
&&-\frac{p}{2n(n-p)}\bigg(i_{X_a}(P_b\wedge i_{X^b}\omega)+i_{X_a}(R_{bc}\wedge i_{X^b}i_{X^c}\omega)\bigg)\nonumber\\
&=&0.
\end{eqnarray}
In general, this is a very restrictive condition on the curvature characteristics of the manifold and gauge curvature $F$ related to the ordinary CKY forms of the background. However, a simplification in (52) occurs if we consider the constant curvature manifolds. In this case, the curvature 2-forms are written as $R_{ab}=\frac{\cal{R}}{n(n-1)}e_a\wedge e_b$ and Ricci 1-forms are $P_a=\frac{\cal{R}}{n}e_a$ while $K_a=-\frac{\cal{R}}{2n(n-1)}e_a$. By substituting them in (52) and using the equality $e^a\wedge i_{X_a}\omega=p\omega$, one can see that the terms that contain curvature characteristics (the terms that does not contain $F$) cancel each other and only the terms that contain $F$ remain. Moreover, constant curvature manifolds are conformally-flat, namely the conformal 2-forms defined in (18) are equal to zero $C_{ab}=0$. In Lorentzian space-times, the gauge curvature $F$ can be determined from $C_{ab}$ by using the integrability condition (19) of the gauged twistor equation \cite{Cassani Martelli}. Indeed, in constant curvature manifolds, the gauge curvature $F=0$ with non-zero $A$, so we have flat connections in the definition of the gauged covariant derivative \cite{Cassani Martelli, de Medeiros2}. This means that the remaining terms that contain $F$ in (52) will be equal to zero and the vanisihing condition of the left hand side of (52) is satisfied. Hence, the terms that are in the coefficients of $\psi$ satisfy (37).

In that way, we prove that the operator defined in (35) is a symmetry operator for the gauged twistor equation in constant curvature manifolds. So, one can construct gauged twistor spinors which are the supersymmetry generators of supersymmetric field theories coupled to supergravity from a known solution by using the ordinary CKY forms of the constant curvature background through (35). The constant curvature manifolds such as anti-de Sitter (AdS) space-times are important since they occur in the backgrounds of supergravity theories and supersymmetric field theories.

On the other hand, CKY forms and gauged twistor spinors can still exist in non-constant curvature manifolds. However, the construction of symmetry operators is more complicated in that case. In some algebraically special spacetimes such as having Petrov type II, III or D, the gauge curvature $F$ can have some zero components although not totally zero \cite{de Medeiros2}. By considering the components of the curvature characteristics and the gauge curvature, one can still find some CKY forms that satisfy (52) to construct the symmetry operators of gauged twistor spinors in more general cases. 

\subsection{From ordinary twistors to gauged twistors}

The construction of the symmetry operator (35) gives rise to a relation between ordinary twistor spinors and gauged twistor spinors. The spinor bilinears of ordinary twistor spinors correspond to the ordinary CKY forms \cite{Acik Ertem}. This means that the symmetry operators in (35) can be written in terms of the ordinary twistor spinors. For an ordinary twistor spinor $\phi$, the $p$-form Dirac currents as defined in (21)
\begin{equation}
(\phi\overline{\phi})_p=(\phi, e_{a_p...a_2a_1}.\phi)e^{a_1a_2...a_p}
\end{equation}
can be replaced with the CKY $p$-forms in the symmetry operator (35). So, the symmetry operator of a gauged twistor spinor $\psi$ can be constructed from the $p$-form Dirac currents of ordinary twistor spinors as
\begin{equation}
L_{\phi\overline{\phi}}\psi=-(-1)^p\frac{p}{n}(\phi\overline{\phi})_p.\widehat{\displaystyle{\not}D}\psi+\frac{p}{2(p+1)}d(\phi\overline{\phi})_p.\psi+\frac{p}{2(n-p+1)}\delta(\phi\overline{\phi})_p.\psi
\end{equation}
and this means that the ordinary twistor spinors generate the solutions of the gauged twistor equation. The exterior derivative and co-derivative of the $p$-form Dirac currents of the ordinary twistor spinors can be found as \cite{Acik Ertem}
\begin{equation}
d(\phi\overline{\phi})_p=\frac{p+1}{n}\bigg(\displaystyle{\not}d(\phi\overline{\phi})-2i_{X^a}(\phi\overline{\nabla_{X_a}\phi})\bigg)_{p+1}
\end{equation}
\begin{equation}
\delta(\phi\overline{\phi})_p=-\frac{n-p+1}{n}\bigg(\displaystyle{\not}d(\phi\overline{\phi})-2e^a\wedge(\phi\overline{\nabla_{X_a}\phi})\bigg)_{p-1}
\end{equation}
and (54) can also be written in the following form
\begin{eqnarray}
L_{\phi\overline{\phi}}\psi&=&-\frac{p}{n}\bigg[(-1)^p(\phi\overline{\phi})_p.\widehat{\displaystyle{\not}D}\psi+\frac{1}{n}\bigg(\big(i_{X^a}(\phi\overline{e_a.\displaystyle{\not}D\phi})\big)_{p+1}-\big(e^a\wedge(\phi\overline{e_a.\displaystyle{\not}D\phi})\big)_{p-1}\bigg).\psi\bigg]\nonumber\\
&&+\frac{p}{2n}\bigg[\big(\displaystyle{\not}d(\phi\overline{\phi})\big)_{p+1}-\big(\displaystyle{\not}d(\phi\overline{\phi})\big)_{p-1}\bigg].\psi
\end{eqnarray}
For $A=0$, this reduces to the symmetry operators of ordinary twistor spinors written in terms of ordinary twistor spinors \cite{Ertem2}.

The symmetry operators of the gauged twistor equation is defined in constant curvature manifolds and the set of CKY forms in that case is of maximal dimension. The maximum number of CKY $p$-forms in $n$ dimensions is \cite{Semmelmann}
\begin{equation}
C_p=\left(
             \begin{array}{c}
               n \\
               p-1 \\
             \end{array}
           \right)+2\left(
                      \begin{array}{c}
                        n \\
                        p \\
                      \end{array}
                    \right)+\left(
                              \begin{array}{c}
                                n \\
                                p+1 \\
                              \end{array}
                            \right)
\end{equation}
and the dimension of the space of ordinary twistor spinors is given in $n$ dimensional constant curvature manifolds as \cite{Lichnerowicz}
\begin{equation}
t=2^{\lfloor n/2\rfloor}+1.
\end{equation}
Those sets of CKY forms or ordinary twistor spinors generate the gauged twistor spinors through (35) or (54).

As an example, let us consider the AdS$_4$ spacetime. The explicit forms of geometric Killing spinors which are twistor spinors corresponding to the eigenspinors of the Dirac operator are calculated in \cite{OFarrill Gutowski Sabra} for AdS$_4$. For the coordinates $t, x, r, \rho$ and a relevant coframe basis ${e^a}$ with $a=0,...,3$, the spinor
\[
\kappa=e^{it}\left(\cosh\rho+i\sinh\rho\right)1-e^{it}\left(\sinh\rho+i\cosh\rho\right)e^2
\]
is a geometric Killing spinor. One can construct more general twistor spinors from geometric Killing spinors by using the Killing reversal method \cite{Fujii Yamagishi, Acik}. So, the following spinor
\[
\phi=\kappa+z.\kappa
\]
is a twistor spinor in AdS$_4$ where $z$ is the volume form. The properties of the spinor inner product in four dimensional Lorentzian manifolds gives the fact that the only non-zero spinor bilinears correspond to 1-form and 2-form Dirac currents \cite{Alexeevsky et al}. So, we can construct the following CKY 1- and 2-forms from the twistor spinor $\phi$
\begin{eqnarray}
\omega_1&=&(\phi, e_a.\phi)e^a\nonumber\\
\omega_2&=&(\phi, e_{ba}.\phi)e^{ab}\nonumber.
\end{eqnarray}
Then, we have the symmetry operators written in terms of CKY forms $\omega_1$ and $\omega_2$ as
\[
L_{\omega_i}=-(-1)^p\frac{p}{4}\omega_i.\displaystyle{\not}{\widehat{D}}+\frac{p}{2(p+1)}d\omega_i+\frac{p}{2(4-p+1)}\delta\omega_i
\]
where $p=1,2$ for $i=1,2$. This operators transforms solutions of the gauged twistor equation to other solutions of it in AdS$_4$. So, one can investigate the algebra structure of these symmetry operators constructed from all twistor spinors in AdS$_4$ to find a mutually commuting set and construct a general solution for the gauged twistor equation, or one can use it to find gauged twistor spinors from the known ones.

\section{Conclusion}

Symmetry operators of the gauged twistor equation in terms of ordinary CKY forms is constructed in constant curvature backgrounds. Since the existence of CKY forms or gauged twistor spinors is a restrictive condition on the underlying manifold, the construction of symmetry operators is constrained to constant curvature manifolds. This is expected, because of the fact that the constant curvature manifolds have maximum numbers of CKY forms and ordinary twistor spinors and constructing symmetry operators using them in constant curvature manifolds is more possible than the other cases. Construction of those symmetry operators provides a new way to obtain the supersymmetry generators of supersymmetric and superconformal field theories in curved backgrounds.

The spinor bilinears of gauged twistor spinors correspond to gauged CKY forms. However, the symmetry operators contain ordinary CKY forms and not gauged CKY forms. This means that the extended superalgebras that contain gauged twistor spinors and gauged CKY forms cannot be obtained by using the constructed symmetry operators while they can be constructed for the cases of ordinary twistor spinors and geometric Killing spinors \cite{Ertem1, Ertem2}. So, one can search for the other types of symmetry operators of gauged twistor equation constructed out of gauged CKY forms and try to construct extended superalgebras from them. These superalgebra structures are important in the classification problem of supergravity and supersymmetric field theory backgrounds.

The methods described in the paper can be used for the explicit constructions of the symmetry operators of gauged twistor spinors in various constant curvature backgrounds. In that way, one can obtain the supersymmetry generators of supersymmetric field theories in those backgrounds. By investigating the algebra structure of those symmetry operators, one can also obtain a commuting set of symmetry operators which provide a general solution for the gauged twistor equation. Moreover, the extensions of the procedures for the construction of the symmetry operators to the cases of more general twistor equations is possible. For example, in the presence of supergravity fluxes, the supergravity twistor equations are coupled to these fluxes and the symmetry operators of those twistor spinors can be investigated in a similar way. 



\begin{references}

\bibitem{Festuccia Seiberg} G. Festuccia and N. Seiberg, "Rigid supersymmetric theories in curved superspace," J. High Energy Phys. JHEP\textbf{1106}, 114 (2011).

\bibitem{Klare Tomasiello Zaffaroni} C. Klare, A. Tomasiello and A. Zaffaroni, "Supersymmetry on curved spaces and holography," J. High Energy Phys. JHEP\textbf{1208}, 061 (2012).

\bibitem{Hristov Tomasiello Zaffaroni} K. Hristov, A. Tomasiello and A. Zaffaroni, "Supersymmetry on three-dimensional Lorentzian curved spaces and black hole holography," J. High Energy Phys. JHEP\textbf{1305}, 057 (2013).

\bibitem{Cassani Martelli} D. Cassani and D. Martelli, "Supersymmetry on curved spaces and superconformal anomalies," J. High Energy Phys. JHEP\textbf{1310}, 025 (2013).

\bibitem{Cassani Klare Martelli Tomasiello Zaffaroni} D. Cassani, C. Klare, D. Martelli, A. Tomasiello and A. Zaffaroni, "Supersymmetry in Lorentzian curved spaces and holography," Commun. Math. Phys. \textbf{327}, 577 (2014).

\bibitem{de Medeiros1} P. de Medeiros, "Rigid supersymmetry, conformal coupling and twistor spinors," J. High Energy Phys. JHEP\textbf{1409}, 032 (2014).

\bibitem{de Medeiros2} P. de Medeiros, "Submaximal conformal symmetry superalgebras for Lorentzian manifolds of low dimension," J. High Energy Phys. JHEP\textbf{1602}, 008 (2016).

\bibitem{Lischewski} A. Lischewski, "Charged conformal Killing spinors," J. Math. Phys. \textbf{56}, 013510 (2015).

\bibitem{Miller} W. Miller, \emph{Symmetry and Separation of Variables} (Addison-Wesley, Massachusetts, 1977).

\bibitem{Benn Charlton} I. M. Benn and P. Charlton, "Dirac symmetry operators from conformal Killing-Yano tensors", Class. Quantum Grav. \textbf{14}, 1037 (1997).

\bibitem{Benn Kress} I. M. Benn and J. Kress, "First-order Dirac symmetry operators", Class. Quantum Grav. \textbf{21}, 427 (2004).

\bibitem{Acik Ertem Onder Vercin} \"{O}. A\c{c}{\i}k, \"{U}. Ertem, M. \"{O}nder and A. Ver\c{c}in, "First-order symmetries of the Dirac equation in a curved background: a unified dynamical symmetry condition", Class. Quantum Grav. \textbf{26}, 075001 (2009).

\bibitem{Houri Kubiznak Warnick Yasui} T. Houri, D. Kubiznak, C. Warnick and Y. Yasui, "Symmetries of the Dirac operator with skew-symmetric torsion," Class. Quantum Grav. \textbf{27}, 185019 (2010).

\bibitem{Kubiznak Warnick Krtous} D. Kubiznak, C. Warnick and P. Krtous, "Hidden symmetry in the presence of fluxes," Nucl. Phys. B \textbf{844}, 185 (2011).

\bibitem{Cariglia Krtous Kubiznak} M. Cariglia, P. Krtous and D. Kubiznak, "Commuting symmetry operators of the Dirac equation, Killing-Yano and Schouten-Nijenhuis brackets,"  Phys. Rev. D \textbf{84}, 024004 (2011).

\bibitem{Cariglia} M. Cariglia, "Hidden symmetries of dynamics in classical and quantum physics," Rev. Mod. Phys. \textbf{86}, 1283 (2014).

\bibitem{Ertem1} \"{U}. Ertem, "Symmetry operators of Killing spinors and superalgebras in $AdS_5$,", J. Math. Phys. \textbf{57} 042502 (2016).

\bibitem{Ertem2} \"{U}. Ertem, "Twistor spinors and extended conformal superalgebras," e-print arXiv:1605.03361 [hep-th] (2016).

\bibitem{Benn Tucker} I. M. Benn and R. W. Tucker, \emph{An Introduction to Spinors and Geometry with Applications in Physics} (IOP Publishing, Bristol, 1987).

\bibitem{Charlton} P. Charlton, "The Geometry of Pure Spinors, with Applications," PhD Thesis (University of Newcastle, 1997).

\bibitem{Baum Friedrich Grunewald Kath} H. Baum, T. Friedrich, R. Grunewald and I. Kath, \emph{Twistors and Killing Spinors on Riemannian Manifolds} (Teubner, Leipzig, 1991)

\bibitem{Baum Leitner} H. Baum and F. Leitner, "The twistor equation in Lorentzian spin geometry," Math. Z. \textbf{247}, 795 (2004)

\bibitem{Benn Kress2} I. M. Benn and J. Kress, "Differential forms relating twistors to Dirac fields," \emph{Differential Geometry and its Applications, Proceedings of the 10th International Conference DGA 2007} pp. 573 (World Scientific Publishing, Singapore, 2008).

\bibitem{Acik Ertem} \"{O}. A\c{c}{\i}k and \"{U}. Ertem, "Higher-degree Dirac currents of twistor and Killing spinors in supergravity theories," Class. Quantum Grav. \textbf{32}, 175007 (2015).

\bibitem{Semmelmann} U. Semmelmann, "Conformal Killing forms on Riemannian manifolds," Math. Z. \textbf{245}, 503 (2003).

\bibitem{Ertem3} \"{U}. Ertem, "Lie algebra of conformal Killing-Yano forms," Class. Quantum Grav. \textbf{33}, 125033 (2016).

\bibitem{Lichnerowicz} A. Lichnerowicz, "On the twistor spinors," Lett. Math. Phys. \textbf{18}, 333 (1989).

\bibitem{OFarrill Gutowski Sabra} J. Figueroa-O'Farrill, J. Gutowski and W. Sabra, "The return of the four- and five-dimensional preons," Class. Quantum Grav. \textbf{24}, 4429 (2007).

\bibitem{Fujii Yamagishi} Y. Fujii and K. Yamagishi, "Killing spinors on spheres and hyperbolic manifolds," J. Math. Phys. \textbf{27}, 979 (1986).

\bibitem{Acik} \"{O}. A\c{c}{\i}k, "Twistors from Killing spinors alias radiation from pair annihilation I: Theoretical considerations," e-print arXiv:1701.05429 [math-ph] (2017).

\bibitem{Alexeevsky et al} D. V Alexeevsky, V. Cortes, C. Devchand and A. Van Proeyen, "Polyvector super-Poincar\'{e} algebras," Commun. Math. Phys. \textbf{253}, 385 (2005).

\end{references}
\end{document}